\documentclass[aps,prl,twocolumn,groupedaddress,showpacs]{revtex4}

\usepackage{amsmath,amsthm,amssymb}
\usepackage{graphicx}
\usepackage{psfrag}

\begin{document}

\title{Surface melting of the vortex lattice}

\author{A.\ De\ Col$^1$,  G.I.\ Menon$^2$, 
 V.B.\  Geshkenbein$^{1,3}$, and G.\ Blatter$^1$}

\affiliation{$^1$Theoretische Physik, ETH-H\"onggerberg, 
   CH-8093 Z\"urich, Switzerland}
\affiliation{$^2$The Institute of Mathematical Sciences, 
   C.I.T. Campus, Taramani, Chennai 600\ 113, India}
\affiliation{$^3$L.D.\ Landau Institute for Theoretical 
   Physics, RAS, 117940 Moscow, Russia}

\date{\today}

\begin{abstract}

We discuss the effect of an $(ab)$-surface on the melting transition of
the pancake-vortex lattice in a layered superconductor within a density
functional theory approach.  Both discontinuous and continuous surface
melting are predicted for this system, although the latter scenario
occupies the major part of the low-field phase diagram.  The formation
of a quasi-liquid layer below the bulk melting temperature inhibits the
appearance of a superheated solid phase, yielding an asymmetric
hysteretic behavior which has been seen in experiments. 

\end{abstract}


\maketitle

The transition between vortex solid and vortex liquid phases observed in
the mixed phase of high-$T_c$ superconductors \cite{nelson88} has
generated renewed interest in the problem of melting. In common with
other discontinuous phase transitions, melting should involve the
appearance of both metastable undercooled liquid and overheated solid
phases. Hence, hysteretic behavior is expected upon cycling through the
transition. However, experiments on the layered, high-$T_c$
superconductor BiSSCO \cite{soibel00} reveal an asymmetric hysteresis,
characterized by the appearance of only the supercooled liquid and no
overheated solid. Similar behavior is displayed by ordinary crystals
\cite{Frenken1985,Tartaglino2005}, where such asymmetry is understood to
be a consequence of surface {(pre-)}\break melting: surfaces act as
nucleation centers for the liquid, thereby inhibiting the metastable
solid above the melting transition. However, such surface melting is not
generic and there are experimental systems where the surface remains
solid up to the bulk melting transition \cite{Tartaglino2005}. In this
letter, we study the effects of an ($ab$)-surface on vortex lattice
melting, showing that as the strength of the magnetic field is varied,
the {\it same} surface may exhibit {\it either} `surface non-melting' or
`surface melting' behavior.  The latter scenario applies to the major
part of the low-field phase diagram, in agreement with experiments
\cite{soibel00}. 

Early studies \cite{pietronero79} of simple crystals have focused on the
solid phase and have demonstrated that the surface turns unstable before
the bulk melts. Going beyond such a stability analysis is a difficult
task, as a theory is required which describes both solid and liquid
phases in a unified manner. Qualitative insight can be gained from a
Landau theory \cite{lipowsky83} by including the destabilizing effect of
the surface: two melting scenarios are found, surface melting ($O_2$)
with a continuous- and surface non-melting ($O_1$) with a discontinuous
vanishing of the order parameter at the surface. More elaborate {\it ab
initio} calculations reduce the problem to a mean-field order-parameter
theory and determine the free energy either as a lattice sum
\cite{Trayanov1987} or within a density functional theory (DFT)
\cite{Ohnesorge1994} exploiting liquid-state correlations. 
\begin{figure}[t]
\centering
   \includegraphics [width=8.0 cm] {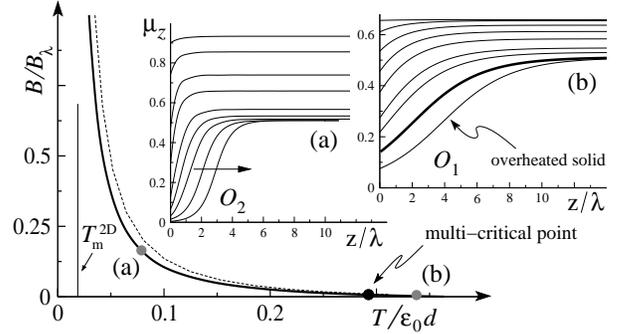}
   \caption{Comparison of the melting line $B_\mathrm{m}(T)$
   obtained via the DFT-substrate approach (present work, full 
   line) with that of Ref.\ \cite{dodgson00} (dashed line).
   Top right: Numerical solutions of the order-parameter profile 
   $\mu_z$ for different fields $B$ increasing from top to bottom: 
   (a) at $T = 0.08 \, \varepsilon_0d$ ($O_2$ transition with
   the liquid-solid interface invading the bulk as $B \nearrow 
   B_\mathrm{m}$) and (b) at $T = 0.33 \, \varepsilon_0d$ 
   ($O_1$ transition with the thick line corresponding to 
   $B = B_\mathrm{m}$).}
   \label{fig:dftbulk}
\end{figure}

The vortex matter in Bi- and Tl- based high-$T_c$ superconductors is
characterized by an extreme anisotropy:  pancake-vortices confined to
superconducting layers exhibit repulsive logarithmic interactions within
the planes and weakly attractive inter-layer interactions extending over
many layers (see, e.g.\ Ref.\ \cite{giannireview};  we ignore here a
weak Josephson coupling between the layers). These anisotropic
properties inspire the use of a substrate--mean-field theory
\cite{dodgson00} describing the vortex system in terms of
two-dimensional lattices of pancake-vortices subject to a substrate
potential generated by the mean action of the vortices residing in other
layers. The bulk melting line is known
\cite{sengupta91,giannireview,dodgson00,fangohr03} to interpolate
between the Berezinskii-Kosterlitz-Thouless (BKT) transition temperature
$T_{\rm\scriptscriptstyle BKT} \lesssim T_c$ of the individual layers at
zero external field and the two-dimensional vortex-lattice melting
temperature $T_\mathrm{m}^{\rm\scriptscriptstyle 2D} \ll T_c$ at high
magnetic fields, see Fig.\ \ref{fig:dftbulk}.  We use the classical DFT
of freezing \cite{ramakrishnan79} and exploit the anisotropic properties
of the system with its natural separation of liquid-state correlations
into in-plane and out-of-plane components.  As a result, we obtain a
reliable and analytically tractable order-parameter theory which allows
us to study $(ab)$-surface melting. We find two regimes within the
$B$-$T$ phase diagram: for low and high magnetic fields $B$ the surface
order parameter undergoes a finite, although reduced, jump ($O_1$) at
the bulk melting temperature, whereas for intermediate values of $B$,
stray magnetic fields destabilize the layers close to the surface,
leading to the surface melting scenario ($O_2$). 

We sketch the derivation of pancake-vortex interactions in a
semi-infinite superconductor filling the half-space $z \ge 0$. These
interactions are mediated by currents set up by the vortices via the
Lorentz force (here, $d$ is the layer separation, $\lambda$ the London
penetration depth, and $\Phi_0$ the flux unit): each vortex generates a
current density ${\bf j} = -(c/4\pi \lambda^2) [(\Phi_0/2\pi)\nabla
\varphi +{\bf A}]$ which acts on another vortex core with a transverse
force ${\bf F} = d \Phi_0\, {\bf z} \times{\bf j}/c$. For a co-planar
pair of pancake-vortices the current density is driven by the phase
gradient $\nabla \varphi= -{\bf n}_z \times {\bf R}/R^2$ and the force
$\propto 1/R$ produces a long-range logarithmic repulsion $V_{z,z}(R)
\approx - 2 \varepsilon_0 d \ln({R}/{\xi})$; this potential corresponds
to that of a one-component charged plasma (OCP) with charge $e^2
\rightarrow 2\varepsilon_0 d$ and $\varepsilon_0 = (\Phi_0/4\pi
\lambda)^2$ the vortex line energy. The interaction between two
pancake-vortices residing in different layers derives from the vector
potential ${\bf A}$; calculating the field associated with a
pancake-vortex in a semi-infinite geometry and integrating the Lorentz
force provides the potential
\begin{eqnarray}
   \label{vzz1}
   &&V_{z,z'}(R) = -\varepsilon_0 d^2
   \int_0^{\infty} \!\!\!\!\! dK
   \frac{J_0(KR)}{\lambda^2 K K_{\scriptscriptstyle +}}
   \\ \nonumber
   && \qquad\qquad\qquad \times [f_{z-z'}(K)+\beta(K) f_{z+z'}(K)]
\end{eqnarray}
with $f_{z}(K) = \exp(-K_{\scriptscriptstyle +} |z|)$,
$K_{\scriptscriptstyle +} = \sqrt{K^2 +\lambda^{-2}}$, and $\beta(K) =
(K_{\scriptscriptstyle +}-K)/ (K_{\scriptscriptstyle +} +K)$. The bulk
term $\propto f_{z-z'}$ is augmented by a stray-field term $\propto
f_{z+z'}$ relevant within a distance $\lambda$ from the surface. For
small in-plane distances $R\ll \lambda$, the contribution of the
stray-field term can be neglected and we recover the bulk expression
$V_{z,z'}(R) \approx$ $\varepsilon_0 d (dR/\lambda^2)[R/(R +4|z-z'|)]$
\cite{giannireview}. For a large separation $R\gg \lambda$, the surface
term is relevant and we obtain
\[
   V_{z,z'}(R) \approx 2 \varepsilon_0 d\bigl[ 
   \phi_\mathrm{t}(z,z')\ln (R/\lambda) + 
   (d/R) e^{-(z+z')/\lambda}\bigr],
\]
where $\phi_\mathrm{t}(z,z')=(d/2\lambda) (e^{-|z-z'|/\lambda}
+e^{-(z+z')/\lambda}) \ll 1$ is the fraction of flux trapped in the
layer $z'$ generated by a pancake-vortex at $z$: a test vortex at $z'$
then effectively experiences a logarithmic attraction from {\it two}
bulk-type pancake-vortices, the real one at $z$ and a fake mirror vortex
with equal sign at $-z$, the latter generated by the stray field.  The
algebraic repulsion associated with the second term in $V_{z,z'}(R)$ is
again due to the stray field and produces a surface softening. 

In our investigation of the vortex solid-liquid transition we make use
of the classical density functional theory (DFT) \cite{ramakrishnan79}
which builds on the (grand canonical) free-energy difference $\delta
\Omega = \Omega_\mathrm{sol} - \Omega_\mathrm{liq}$ expressed through
the variation $\delta \rho({\bf r}) = \rho({\bf r}) - \bar\rho$ in
particle density away from the uniform liquid state density $\bar\rho$; 
for the inhomogeneous and anisotropic vortex matter system, $\delta
\rho_z({\bf R}) = \rho_z({\bf R}) - \bar\rho$ and $\bar\rho$ is the 2D
liquid density,
\begin{eqnarray}
   \lefteqn{\frac{\delta \Omega [\rho_z({\bf R})]} {T}\!
   = \int\! \frac{dz}{d} d^2{\bf R} \, \Bigl[
   \rho_{z}({\bf R})\ln\frac{\rho_{z}({\bf R})}{\bar{\rho}}
   -\delta \rho_{z} ({\bf R})}
   &
   \nonumber\\
   && -\frac{1}{2}
   \int\!\! \frac{dz '}{d} d^2{\bf R'}
   \delta \rho_{z}({\bf R})
   c_{z,z'}(|{\bf R}\!-\!{\bf R}'|)
   \delta \rho_{z'}({\bf R}')\Bigr]. \label{func}
\end{eqnarray}
The first two terms describe the entropic contribution of a
non-interacting gas, while the last term accounts for the microscopic
interactions $V_{z,z'}(R)$ via the direct pair-correlation function
$c_{z,z'}(R)$ of the liquid state; in the homogeneous liquid, the
(dimensionless) Fourier transform $c_{k_z}(K) = (\bar\rho/d) \int d^3r
\, c_z(R) e^{-i{\bf k}\cdot{\bf r}}$ is related to the static structure
factor via $S({\bf k})=1/[1-c_{k_z} (K)]$. The appearance of finite
density-modulations $|\delta \rho_{z}({\bf R})| > 0$ in the solid is a
consequence of these correlations. We exploit the anisotropy of the
pancake-vortex system and separate $c_{z,z'}(K)$ into in- and
out-of-plane parts,
\begin{equation}\label{csub}
  c_{z,z'}(K) =
  c^{\rm\scriptscriptstyle 2D}(K) d \, \delta(z\!-\!z')
   - V_{z,z'}(K)/T,
\end{equation}
where $c^{\rm\scriptscriptstyle 2D}(K)$ denotes the correlation function
of the 2D-OCP as obtained from standard Monte-Carlo simulations
\cite{alvise_long}. The second term accounts for the weak inter-plane
interactions; within lowest-order perturbation theory \cite{hansenbook}
it is given by the Fourier transform of the out-of-plane interaction
(\ref{vzz1}), $V_{z,z'}(K)/T= -\alpha(K) [f_{z-z'}(K) +\beta(K)
f_{z+z'}(K)]$ with $\alpha(K) = 2\pi \varepsilon_0 d^2 \bar\rho/ T
\lambda^2 K^2 K_{\scriptscriptstyle +}$. 

We begin with the homogeneous bulk system. Consider the Fourier
transforms $\delta \rho_z({\bf K_n}) /\bar{\rho}$ for the density field
and $\xi_z({\bf K_n})$ for the molecular field \cite{ramakrishnan79}
$\xi_z({\bf R}) \equiv \ln [\rho_z({\bf R})/\bar{\rho}]$, where ${\bf
K_n}$ denote the reciprocal triangular-lattice vectors. For a
correlation function $c^{\rm \scriptscriptstyle 2D}(K)$ decaying rapidly
beyond the first reciprocal lattice vector, only the first two
components need be retained and we can restrict the Ansatz to the form
$\delta \rho_z({\bf R}) /\bar{\rho} \approx \eta_z +\mu_z g({\bf R})$
and $\xi_z({\bf R}) \approx \kappa_{z}+\xi_z g({\bf R})$ with $g({\bf
R}) = \sum_{|{\bf K_n}| = G} \exp(i{\bf K_n}\cdot {\bf R})$, $G=4\pi
/\sqrt{3} a_{\scriptscriptstyle \triangle}$ ($a_{\scriptscriptstyle
\triangle}$ the lattice constant, $a_{\scriptscriptstyle \triangle}^2 =
2\Phi_0/\sqrt{3}B$). Furthermore, $\kappa_z$ and $\xi_z$ are related to
$\mu_z$ via \cite{explain}
\begin{equation}
   \kappa_z = -\Phi(\xi_z),\quad
   \mu_z = \Phi^\prime(\xi_z)/6,
   \label{rel}
\end{equation}
with $\Phi(\xi) = \ln s^{\scriptscriptstyle -1}\!\! \int_{s} 
d^2 R \exp[\xi g(R)]$ and $\Phi^\prime$
its derivative (here, $s$ denotes the 2D unit cell area). Inserting this
Ansatz into the free-energy density (\ref{func}), we obtain the reduced
functional for a bulk system in the form (we use the normalization
$\delta \omega = \delta \Omega/ T S \bar\rho$ with $S$ the sample area)
\[
   \frac{\delta \omega[\mu_z]}{T}
    \! = \! \int \! \frac{dz}{d} \Bigl[
   \frac{\delta \omega^{\rm\scriptscriptstyle 2D}_\mathrm{sub}
   (\mu_z)}{T} + \frac{3\bar\alpha}{2} \! \int \! \frac{dz'}{d}
   \bar{f}_{z-z'}(\mu_z-\mu_{z'})^2\Bigr],
\]
with $\bar\alpha = \alpha(G)$ and $\bar{f}_{z-z'} = {f}_{z-z'}(G)$. 
The first term describes the free-energy density of individual 
layers 
\begin{equation}
   \frac{\delta \omega^{\rm\scriptscriptstyle 2D}_\mathrm{sub}
   (\mu_z)} {T} = \kappa_z(\mu_z)+6\xi_z(\mu_z)\mu_z-3 \Bigl[\bar{c}^{\rm
   \scriptscriptstyle 2D}+\frac{2 \bar\alpha}
   {d\,G_{\scriptscriptstyle +}}\Bigr]\mu_z^2,
   \label{free2d}
\end{equation}
with in-plane correlations $\bar{c}^{\rm \scriptscriptstyle 2D} \equiv
c^{\rm \scriptscriptstyle 2D}(G)$ and out-of-plane correlations
described by the substrate potential $-\bar\alpha \int (dz/d)
\bar{f}_{z} = -2 \bar\alpha/d\,G_{\scriptscriptstyle +}$ with
$G_{\scriptscriptstyle +} = \sqrt{G^2+\lambda^{-2}}$.  Note that both
$\kappa_z$ and $\xi_z$ have to be understood as functions of the
order-parameter $\mu_z$ via (\ref{rel}).  The second non-local term in
$\delta\omega$ accounts for inhomogeneities of the order-parameter field
$\mu_z$ and involves the dispersive `elastic coefficient' $\bar\alpha
\bar{f}_{z-z'}$. Here, we neglect the small change in density across the
transition described by $\eta_z$;  its inclusion involves a more
elaborate analysis accounting for the discrete nature of the particles
\cite{alvise_long}. 

\begin{figure}[ht]
   \centering
   \includegraphics [width= 8.0cm] {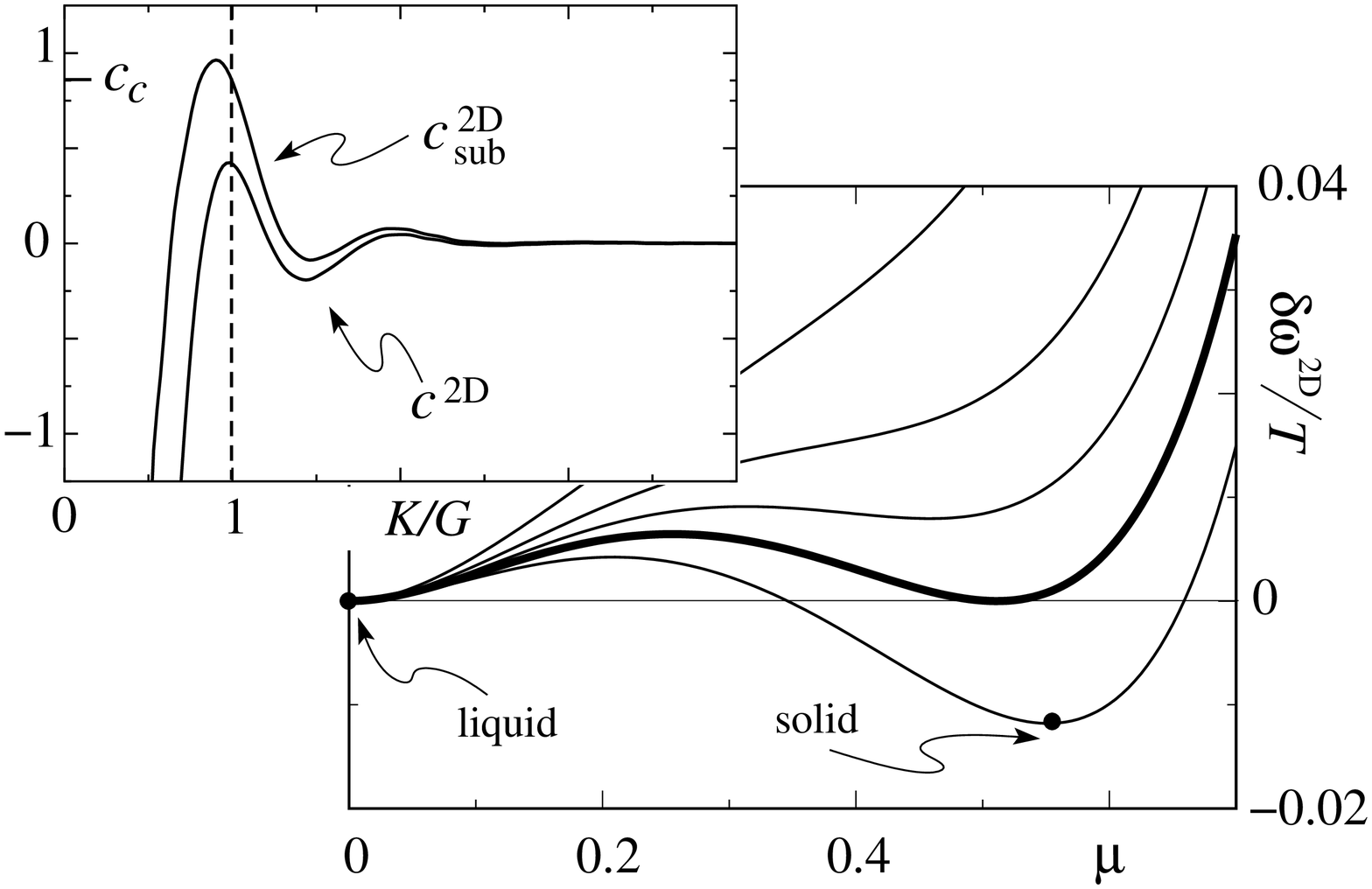}
   \caption{Top left: direct pair-correlation functions at
   $T/ \varepsilon_0 d =0.08$ for the 2D-OCP,
   $c^{\rm \scriptscriptstyle 2D}(K)$ from MC simulations 
   and $c^{\rm\scriptscriptstyle 2D}_\mathrm{sub}(K)=
   c^{\rm \scriptscriptstyle 2D}(K) + 2\alpha(K)/dK_{ \scriptscriptstyle +}$ 
   for the full 3D system at melting ($B = B_\mathrm{m}$). 
   Bottom right: Free energy $\delta \omega^{\rm\scriptscriptstyle 
   2D}(\mu)/T$ for values $\bar{c}^{\rm\scriptscriptstyle 2D}$
   = 0.80, 0.83, 0.845, 0.856 ($=\bar{c}_c$ critical value,
   thick line), 0.87. At melting the order parameter jumps
   from $\mu_\mathrm{sol}\approx 0.51$ to $\mu_\mathrm{liq} =0$.}
   \label{fig:2Denergy}
\end{figure}

The bulk melting line $B_\mathrm{m}(T)$ is obtained from minimizing the
functional $\delta\omega[\mu]$ for a homogeneous order parameter $\mu$.
At large values of $B$, the inter-planar interaction is negligible and
the melting temperature is given by the solid-liquid transition of the
2D-OCP \cite{ramakrishnan79} as described by the free-energy density
(\ref{free2d}) without substrate potential (termed
$\delta\omega^{\rm\scriptscriptstyle 2D}$). For small $\bar{c}^{\rm
\scriptscriptstyle 2D}$ (large temperatures), $\delta
\omega^{\rm\scriptscriptstyle 2D}(\mu)$ exhibits only the liquid minimum
at $\mu_\mathrm{liq} = 0$, cf.\ Fig.\ \ref{fig:2Denergy}. Lowering the
temperature, the correlator $\bar{c}^{\rm\scriptscriptstyle 2D}$
increases and $\delta \omega^{\rm\scriptscriptstyle 2D}(\mu)$ develops a
second minimum at a finite value of $\mu$ describing the solid phase. 
At the critical value $\bar{c}^{\rm\scriptscriptstyle 2D} = \bar{c}_c =
0.856$ the `solid' minimum drops below the `liquid' one and the
high-field liquid-solid transition takes place. The comparison with
numerical simulations \cite{caillol} allows us to check the accuracy of
our approach: Monte Carlo simulations \cite{caillol} show that the
2D-OCP freezes at $T_\mathrm{m}^{\rm \scriptscriptstyle 2D} \approx
\varepsilon_0 d/70$ where the correlator assumes the value $\bar{c}^{\rm
\scriptscriptstyle 2D} \approx 0.77 < \bar{c}_c$; more sophisticated
versions of DFT cure this discrepancy \cite{singh}. 

At low $B$ ($< B_\lambda \equiv \Phi_0/\lambda^2$), the inter-planar
correlations become relevant and the 2D correlations $\bar{c}^{\rm
\scriptscriptstyle 2D}$ are augmented by the substrate potential. The
condition $\bar{c}_c = \bar{c}^{\rm \scriptscriptstyle
2D}+2\bar{\alpha}/ d G_{\scriptscriptstyle +} = \bar{c}^{\rm
\scriptscriptstyle 2D} +(\sqrt{3}\varepsilon_0d/2\pi T) [1+(8\pi^2
/\sqrt{3}) B/B_\lambda]^{-1} $ yields the bulk melting line 
\begin{equation}
   \label{bt_bulk}
   \frac{B_\mathrm{m}(T)}{B_\lambda}=\frac{\sqrt{3}}{8\pi^2} 
   \Bigl[\frac{\sqrt{3}\varepsilon_0 d} {2\pi T [\bar{c}_c
   -\bar{c}^{\rm\scriptscriptstyle 2D}(T)]} -1 \Bigr].
\end{equation}
Given the temperature dependence of the OCP correlator
$\bar{c}^{\rm\scriptscriptstyle 2D}(T)$ as an input, we obtain the full
melting line as shown in Fig.\ \ref{fig:dftbulk}; our result agrees well
with those from numerical simulations \cite{fangohr03} and improves upon
results obtained from other liquid-state closure schemes
\cite{sengupta91}. 

Surface melting involves a non-uniform order-para\-meter field $\mu_z$
defined in a half-infinite sample $z>0$.  The translation-invariant
correlator $\propto \bar{f}_{z-z'} $ in $\delta\omega$ now has to be
replaced by the full expression $V_{z,z'}(G)/T$ with additional mirror
and surface terms. Minimization with respect to $\mu_z$ provides us with
the integral equation
\begin{eqnarray}
   &&\frac{\partial_{\mu_z} \delta \omega^{\rm
   \scriptscriptstyle 2D}_\mathrm{sub}(\mu_z)}{T}
   + 6\bar\alpha\!\! \int_0^\infty\! \frac{dz'}{d}
   [\bar{f}_{z-z'}+\bar{f}_{z+z'}] (\mu_z-\mu_{z'}) \nonumber\\
   &&\qquad\qquad\quad
   + \frac{12\bar\alpha G}{G+G_{\scriptscriptstyle +}} \int_0^\infty\! 
   \frac{dz'}{d}\, \bar{f}_{z+z'}\, \mu_{z'}= 0.
   \label{eqmotion_0}
\end{eqnarray}
The order parameter profile $\mu_z$ can be calculated numerically and we
show two typical examples in Fig.\ \ref{fig:dftbulk}: while we find the
surface order-parameter $\mu_0$ to vanish smoothly at $T= 0.08\,
\varepsilon_0 d$ ($O_2$ scenario), a residual jump in $\mu_0$ is
obtained at the higher temperature $T= 0.33\, \varepsilon_0 d$ ($O_1$).
In the following, we locate the multi-critical point $T_\mathrm{mc}$ on
the melting line $B_\mathrm{m}(T)$ separating the continuous $O_2$- from
the discontinuous $O_1$ regime. 

To make progress analytically, we have to simplify the non-local terms
in (\ref{eqmotion_0}). Concentrating on the bulk, a gradient expansion
of the term $\propto \bar{f}_{z-z'}$ leads us to the differential
equation (we define $\partial_z \mu_z \equiv \mu_z^\prime$)
\begin{equation}
   \ell^2 \mu_z^{\prime\prime}
   = \partial_{\mu_z}\, \delta \omega^{\rm\scriptscriptstyle
   2D}_\mathrm{sub}(\mu_z)/T
   \label{eqmotion}
\end{equation}
with $\ell^2 = (6\bar\alpha/d)\int dz $ $\bar{f}_z z^2 = 12
\bar\alpha/d\, G_{\scriptscriptstyle +}^3$. This local approximation is
valid if the profile $\mu_z$ varies slowly over the extension
$G_{\scriptscriptstyle +}^{-1}$ of the kernel; we have verified that
this condition is fulfilled for $T \gtrsim 0.04\, \varepsilon_0 d$
(corresponding to $B \lesssim 0.5\, B_\lambda$). Equation (\ref{eqmotion})
has to be completed with a boundary condition. Close to $T_\mathrm{mc}$,
the order parameter $\mu_z$ is small near the surface; linearizing 
$\partial_{\mu_z} \delta \omega^{\rm \scriptscriptstyle 2D}_\mathrm{sub}
/T\approx 6 (1-\bar{c}_c)\mu_z$ in (\ref{eqmotion_0}) we obtain 
the equation 
\begin{equation}
   \mu_z = \frac{G_{\scriptscriptstyle +}}{2(1+r)} \int_0^\infty\! dz'\,
   [\bar{f}_{z-z'}+\bar\beta \bar{f}_{z+z'}]\, \mu_{z'}
   \label{inteq}
\end{equation}
with $r=d G_{\scriptscriptstyle +}(1-\bar{c}_c)/2\bar\alpha$ and $\bar
\beta = (G_{\scriptscriptstyle +} -G)/ (G_{\scriptscriptstyle +} +G)$. 
While the solution of this type of integral equation is a non-trivial 
task in general, a straightforward solution is possible in the present 
case due to the particular exponential structure of the kernel. Taking the 
second derivative of (\ref{inteq}), we obtain the differential equation 
$\ell^2 \mu_z^{\prime\prime} = 6(1-\bar{c}_c) \mu_z$, which shows no 
trace of the boundary term $\propto \beta$ and thus coincides with 
the bulk equation (\ref{eqmotion}) at small $\mu_z$ \cite{renorm}.
This equation then admits the exponential solution $A\exp(\sqrt{r} 
G_{\scriptscriptstyle +}z)+B\exp(-\sqrt{r}G_{\scriptscriptstyle +}z)$ 
and inserting this Ansatz back into (\ref{inteq}) the boundary term 
fixes the ratio $A/B$; as a result, we obtain the boundary condition 
\begin{equation}
   [\mu_z^\prime /\mu_z]_{z=0}=G_{\scriptscriptstyle +}
   (1-\beta)/(1+\beta)= G.
   \label{bc}
\end{equation}

\begin{figure}[ht]
   \centering
   \includegraphics [width= 7.2cm] {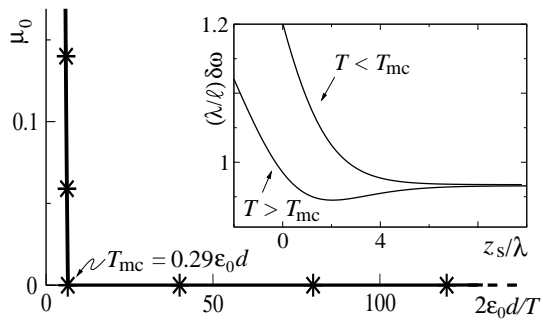}
   \caption{Order parameter $\mu_0$
   at the surface: analytic (see Eq.\ (\ref{surfcond}),
   solid line) and full numerical solution (stars).
   Inset: energy as a function of soliton position
   $z_\mathrm{s}$. Note the minimum for
   $T = 0.33\, \varepsilon_0 d > T_\mathrm{mc}$
   ($O_1$, pinned soliton) which has disappeared at
   $T = 0.28\, \varepsilon_0 d < T_\mathrm{mc}$
   ($O_2$, depinned soliton).}
   \label{fig:mu0}
\end{figure}

The analysis of the boundary value problem (\ref{eqmotion}) 
with (\ref{bc}) follows the one in Ref.\ \cite{lipowsky83}: 
Combining the boundary condition (\ref{bc}) with the expression 
for the `conserved energy' $(\ell\, \mu_z^\prime)^2-2\delta 
\omega^{\rm\scriptscriptstyle 2D}_\mathrm{sub}(\mu_z)/T = 0$, 
we find the relation 
\begin{equation}\label{surfcond}
   \mu_0 \ell G = \sqrt{2\delta\omega^{\rm
   \scriptscriptstyle 2D}_\mathrm{sub}(\mu_0)/T}.
\end{equation}
The liquid surface $\mu_0=0$ is always a solution and we deal with a
continuous surface melting ($O_2$ scenario) if it is the only one. Once
a second solution with $\mu_0>0$ is present, the surface undergoes a
discontinuous $O_1$ transition.  The $O_1$- and $O_2$ scenarios are
separated by a multi-critical point: expanding $\delta \omega^{\rm
\scriptscriptstyle 2D}_\mathrm{sub}$, we find that (\ref{surfcond}) 
admits two solutions for $T> T_\mathrm{mc} \approx 0.29\, 
\varepsilon_0 d$ ($B_\mathrm{mc} \approx 0.007 B_\lambda$) 
and the surface (non)-melting is realized for $T<T_\mathrm{mc}$ 
($T>T_\mathrm{mc}$). In Fig.\ \ref{fig:mu0}, we show the solution of 
(\ref{surfcond}) and compare it with numerical results. 

The appearance of the multi-critical point $(T_\mathrm{mc},
B_\mathrm{mc})$ can be interpreted as a surface-depinning transition of
the solid-liquid interface (soliton): the energy cost to push this
soliton into the system generates a repulsive potential, while the
surface term favoring the liquid phase helps the soliton to enter, see
Fig.\ \ref{fig:mu0}. With decreasing temperature the stable minimum
close to the surface (generating a finite $\mu_0$ and leading to a $O_1$
transition) moves deeper into the bulk and disappears altogether at
$T_\mathrm{mc}$.  As a consequence, the (half-height) position $z_s$ 
of the solitonic solid-liquid interface diverges logarithmically into 
the bulk \cite{lipowsky83} as $T \to T_\mathrm{m}$ from below, 
$z_\mathrm{s} \sim |\ln(T_\mathrm{m}- T)|$, within the entire $O_2$ 
regime $B > B_\mathrm{mc}$. 

Finally, at high magnetic fields the layers melt independently 
following a first-order type 2D melting scenario. The order
parameter $\mu_0$ in the topmost layer then undergoes a finite
jump and the surface non-melting ($O_1$) scenario applies,
implying the existence of a second multi-critical point
at high fields.  Indeed, we do find such a jump at fields
of order $10\, B_\lambda$, however, a more elaborate version of DFT
\cite{singh} is required for an accurate determination of this second
multi-critical point. 

In conclusion, we have analyzed surface melting in the pancake-vortex
system of layered superconductors and have found both surface-melting
($O_2$) and surface--non-melting ($O_1$) scenarios. The $O_2$ scenario
is realized over most of the low-field phase diagram and explains the
experimental observation of asymmetric hysteresis \cite{soibel00}. 

We acknowledge financial support from the Swiss National Foundation
through the program MaNEP and the DST (India).


\begin{thebibliography}{99}
\bibitem{renorm} We drop a small renormalization factor $(1+r) \ell^2
   \mu_z^{\prime\prime} \approx \ell^2 \mu_z^{\prime\prime}$
   which is absent in the gradient expansion of the bulk equation,
   the difference being due to our neglecting of higher-order 
   corrections in $\mu_z^\prime$. The factor $(1+r)$ thus is 
   beyond the accuracy of the full non-linear problem.
\end{thebibliography}
\end{document}